\documentclass[pss,fleqn]{w-art}
\usepackage{times}
\usepackage{w-thm}
\usepackage[]{graphicx}
\begin{document}
\newcommand {\bea}{\begin{eqnarray}}
\newcommand {\eea}{\end{eqnarray}}
\newcommand {\bb}{\bibitem}
\DOIsuffix{theDOIsuffix}
\Volume{XX}
\Issue{1}
\Month{01}
\Year{2003}
\pagespan{1}{}
\Receiveddate{}
\Reviseddate{}
\Accepteddate{}
\Dateposted{}
\title{Resonance impurity scattering in superconductivity in PrOs$_{4}$Sb$_{12}$}


\author[D. Parker]{D. Parker\inst{1}}
\address[\inst{1}]{Department of Physics and Astronomy, University of Southern
California, Los Angeles, CA 90089-0484 USA}
\author[S. Haas]{S. Haas\inst{1}}
\author[K. Maki]{K. Maki\inst{1}}

\begin{abstract}
We study the effect of impurity scattering in the unitarity limit on the
A and B phase superconductivity in PrOs$_{4}$Sb$_{12}$.  We take the triplet
p+h-wave superconducting order parameters and the impurity scattering is
treated within the standard theory.  We find the quasiparticle density of
states and thermodynamics are very sensitive to the impurity scattering.
The impurity scattering dependence of the superconducting transition 
temperature and the superconducting order parameter at $T=0\,K$ are very
similar to those in d-wave superconductors.  Some of these characteristics
will provide a sensitive test of the symmetry of the underlying superconducting
order parameters.
\end{abstract}
\maketitle                   






\section{Introduction}

Superconductivity in the cubic heavy-fermion (HF) filled skutterudite 
PrOs$_4$Sb$_{12}$ was discovered in 2002 
by Bauer et al\cite{bauer,vollmer}, and since that time many experimental and theoretical 
studies of this compound have been performed.  Vollmer et al \cite{vollmer}
studied the low-temperature specific heat of this compound, finding a double
superconducting transition with T$_{c}$'s of 1.75 and 1.85 K.  This material 
thus appears to have at least two thermodynamic phases, which we term the
A and B phases in the following.  Vollmer et al also
found the value of the specific heat jump $\Delta C/C$ at T$_{c}$ to be
3, including both transitions.  

Much effort has centered on establishing the phase diagram of 
PrOs$_{4}$Sb$_{12}$, with several measurements of the upper critical field
H$_{c2}$ \cite{tenya,ho2,tayama}. This material is an extreme 
type 2 superconductor \cite{mclaughlin}, 
so that in the intermediate state between H$_{c1}$ and H$_{c2}$ partial flux
penetration in the form of vortices appears.  All H$_{c2}$ 
measurements indicate an upper critical field near T=0 of 
approximately 2 Tesla.  Only two measurements of the A-B
phase boundary have been made.  Izawa et al
\cite{izawa} performed a measurement of the magnetothermal conductivity of
PrOs$_{4}$Sb$_{12}$ and found that the higher-temperature, higher-field A
phase contains at least four order parameter nodes along the cubic axes, 
and that the lower-temperature, lower-field B phase contains only two nodes
along the [010] axis.  
Measson et al \cite{measson} detected the A-B phase boundary
through thermodynamic measurements of the specific heat (which undergoes a
jump in the transition from one phase to another) and initially 
produced an alternative
phase diagram, in which the A-B phase boundary lies much closer to the upper
critical field H$_{c2}$, and essentially parallels this curve.  
Recent work by Measson, however \cite{sces} details several samples which
appear to contain only an A phase, so that the phase diagram of this material
remains highly controversial.  Below we present two alternative diagrams
based upon the work of Measson et al and Izawa et al.
\begin{figure}[ht]
\includegraphics[width=6cm]{meas.eps}
\includegraphics[width=6cm]{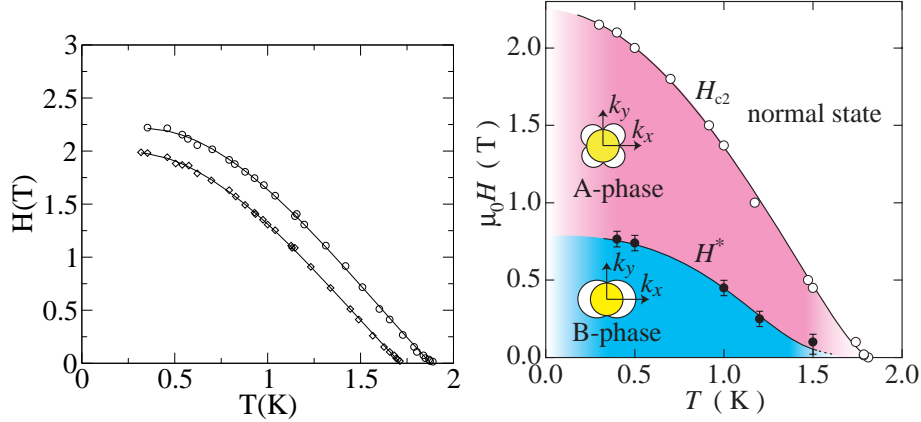}
\caption{Phase diagrams by Measson et al (left) and Izawa et al.  The 
top line and data points represent H$_{c2}$ of the A phase, while the lower
one represents the AB phase boundary H*.}
\end{figure} 

Much effort has also centered on the determination of the 
order parameter symmetry of these phases.
Kotegawa \cite{kotegawa}
investigated the nuclear spin-lattice relaxation time via nuclear quadrupolar
resonance experiments, and found evidence for 
exponentially activated behavior above 
$0.3 T_{c}$, indicating isotropic or s-wave pairing.  Kotegawa also found a
large value of $\Delta/T_{c}$ of approximately 2.5, significantly exceeding 
the BCS s-wave weak-coupling value of 1.76.  Similarly, MacLaughlin et al 
\cite{mclaughlin} studied the pairing state via muon-spin resonance and also 
found evidence for an isotropic gap.  Suderow et al \cite{suderow} performed
scanning tunneling microscopy (STM) measurements and asserted that the Fermi surface
is gapped over most, if not all, quasiparticle momenta.  As a 
directionally locked probe, however, this measurement may have difficulty
determining certain nodal structures, such as point nodes.

Other experiments show strong
evidence for nodal superconductivity in PrOs$_{4}$Sb$_{12}$.  In
particular, Chia et al \cite{chia} measured the superfluid density of 
single crystals and found low-temperature power-law behavior, with
$\rho_{s} \sim 1 - \alpha T^{2}$. This power-law behavior was observed for fields parallel
to each of the three crystal axes.  In addition, the 
previously mentioned magnetothermal
conductivity data of Izawa et al \cite{izawa} strongly suggests the presence
of point nodes.  

Several theoretical proposals have been put forth in an attempt to understand
the superconductivity in PrOs$_{4}$Sb$_{12}$.  Miyake et al \cite{miyake}
proposed triplet p-wave pairing caused by quadrupolar fluctuations, whereas 
Goryo \cite{goryo} proposed an ``anisotropic s-wave'' state for the A phase
and an ``anisotropic s+id-wave'' state for the B phase, 
assuming singlet pairing.  Sergienko and Curnoe
\cite{sergienko} proposed states with a mixture of line nodes and point nodes.
For the purposes of this paper, we will assume the 
existence of unconventional superconductivity in PrOs$_{4}$Sb$_{12}$ with
A-phase and B-phase order parameters as given in Ref. \cite{maki1}, i.e.
triplet p+h-wave order parameters.  The 
clean-limit thermodynamics of these order parameters 
was worked out (as in \cite{salerno}) in 
\cite{parker2}, while the upper critical field H$_{c2}$ was worked
out in \cite{parker3}.  In both works reasonable 
agreement between theory and experiment 
was found, suggesting further analysis.  

Some work concerning superconductivity in substituted samples of PrOs$_{4}$Sb$_{12}$ has
been performed.  Frederick et al \cite{frederick} and Chia et al \cite{chia2}
both studied the superconducting order in single crystals of 
Pr(Os$_{1-x}$Ru$_{x}$)$_{4}$Sb$_{12}$.  Frederick found 
evidence for unconventional
superconductivity in pure PrOs$_{4}$Sb$_{12}$, but observed exponentially
activated behavior in all of the other samples with $x \geq 0.05$.  Chia found
evidence for a three phase scenario at low dopings ($ x < 0.2$), and confirmed 
the exponentially activated behavior found by Frederick.  Rotundu et al
\cite{rotundu} examined the low-temperature susceptibility and specific heat of
single crystals of Pr$_{1-x}$La$_{1-x}$Os$_{4}$Sb$_{12}$ and found that the 
Lanthanum substitution had little effect on the superconductivity, with
an essentially linear decrease in T$_{c}$ as x as increased.  

The recent interest in substituted samples suggests that theoretical modeling of
the effects of impurity scattering is in order.  We expect the theory outlined below,
which assumes strong impurity scattering, should be applicable when samples doped
towards non-superconducting materials (unlike PrRu$_{4}$Sb$_{12}$ and LaOs$_{4}$Sb$_{12}$)
are tested.

\section{Impurity Scattering in PrOs$_{4}$Sb$_{12}$}

In this section we present the results of calculations of the effects of
impurity scattering on the superconductivity in PrOs$_{4}$Sb$_{12}$.  As 
mentioned previously, we use here the order parameter model 
described in Ref. \cite{maki1}
As 
we expect the effects of impurity scattering to be quite strong we 
model the scattering in the unitarity limit, in which the phase shift between
the incoming and scattered waves is taken to be $\pi/2$.  Then following the
approach described in Ref. \cite{g.f.yang} we incorporate 
the scattering by renormalizing
the quasiparticle frequency $\tilde{\omega}$ as follows:
\bea
\tilde{\omega} &=& \omega + i \Gamma {\left\langle\frac{\tilde{\omega}}
{\sqrt{\tilde{\omega}^{2}-\Delta^{2}f^{2}}}\right\rangle}^{-1}
\eea
Here $\Gamma$ is the quasiparticle scattering rate in the normal state and $\Delta=
\Delta(T)$.  For the A-phase $f=\frac{3}{2}
(1-\hat{k}_{x}^{4}-\hat{k}_{y}^{4} -\hat{k}_{z}^{4})$, and 
$\langle \dots \rangle$ denotes $\int d\Omega/4\pi$.
For the B-phase, $f=1-z^{4},$ and $\langle \ldots \rangle$
denotes $\int_{0}^{1}dz \ldots$. 

The quasiparticle density-of-states (DOS) can be obtained once $\tilde{\omega}$
is known, from the relation 
\bea
N(E)&=& N_{0} Re\left\langle\frac{\tilde{\omega}}
{\sqrt{\tilde{\omega}^{2}-\Delta^{2}f^{2}}}\right\rangle
\eea

Below is the density-of-states for the A and B phases for a few impurity
scattering rates, as indicated.  As is evident from the plots, even a small
amount of impurity scattering creates significant low-energy
excitations and truncates the peak normally observed at $E=\Delta$.  It is
interesting that N(E) is largely independent of E for relatively low
energy.  These DOS should be experimentally accessible via scanning 
tunneling microscopy (STM) measurements.
\begin{figure}
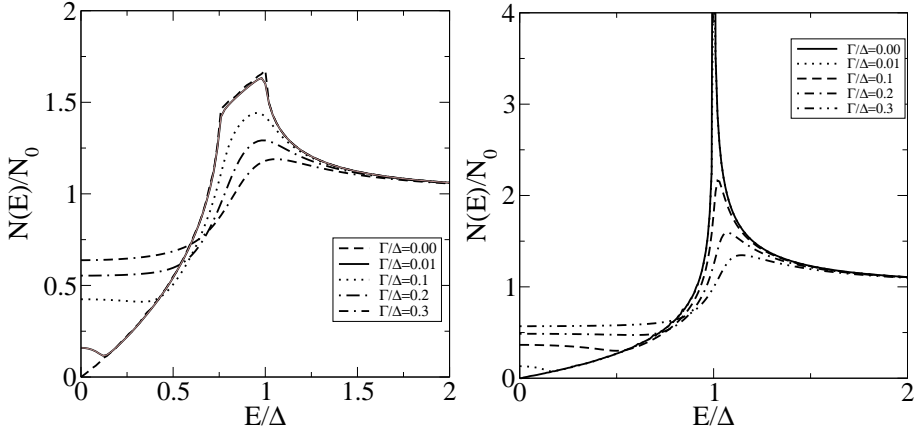

\includegraphics[width=6cm]{dosAplot.eps}
\includegraphics[width=6cm]{dosB.eps}
\caption{The quasiparticle density-of-states for the A phase (left) and
the B phase are shown, for several impurity scattering rates.}
\end{figure}

Also of interest is the overall dependence of the zero-energy DOS on
impurity scattering rates, along with the zero-temperature behavior 
$\Delta(\Gamma,T=0)$ and T$_{c}(\Gamma)$.  These quantities are shown
below for both phases.  As stated previously, the zero-energy DOS increases
quite rapidly for small impurity concentrations.  Here T$_{c}(\Gamma)$
is found by solving the well-known Abrikosov-Gorkov equation \cite{abrikosov}:
\bea
-\ln(T_{c}/T_{c0}) &=& \Psi(\frac{1}{2}+ \frac{\Gamma}{2\pi T_{c}}) - 
\Psi(\frac{1}{2})
\eea
Here $\Psi(z)$ is the digamma function.  And $\Gamma_{c}$, the critical
impurity scattering rate where superconductivity disappears, is given
by 0.8819 T$_{c0}$ for both phases.

\begin{figure}
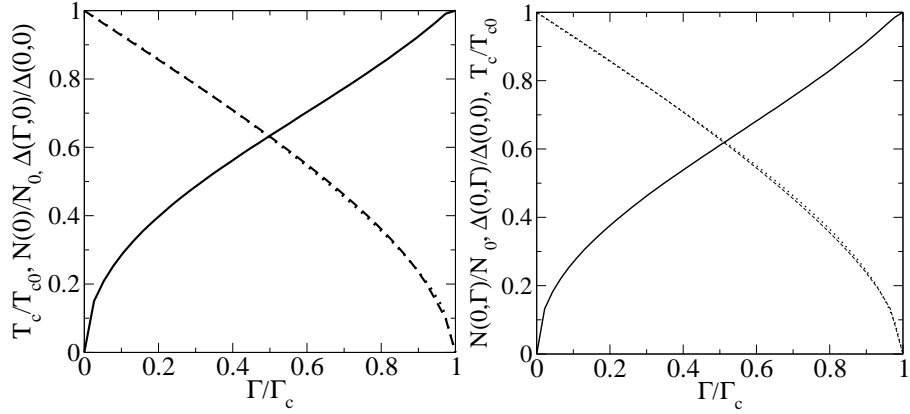

\includegraphics[width=6cm]{AdeltaGamma.eps}
\includegraphics[width=6cm]{Bgamma.eps}
\caption{The zero-energy DOS, zero-temperature $\Delta(\Gamma)$ and
T$_{c}(\Gamma)$ for the A phase (left) and
the B phase are shown.  The zero-energy DOS begins at the lower left corner
of each plot.}
\end{figure}

The temperature-dependent order parameter $\Delta(T)$ is determined by solving the
BCS weak-coupling gap equation:
\bea
\lambda^{-1}& = & 2\pi T<f^{2}>^{-1}\sum_{n}^{}\langle\frac{f^{2}}
{\sqrt{{\tilde{\omega}_{n}}^{2}+\Delta^{2}f^{2}}}\rangle
\eea
Here $\omega_{n}$ and $\tilde{\omega}_{n}$ are related by 
\bea
\tilde{\omega}_{n} &=& \omega_{n} + \Gamma 
{\left\langle\frac{\tilde{\omega}_{n}}
{\sqrt{{\tilde{\omega}_{n}}^{2}+\Delta^{2}f^{2}}}\right\rangle}^{-1}
\eea
Here $\omega_{n}$ is the Matsubara frequency $\omega_{n}= (n+\frac{1}{2})
2\pi T$.  In addition, the sum over n must be truncated at $\omega= E_{0}$, the
cut-off energy.  Plots of the temperature dependent order parameters
$\Delta_{A}(T)$ and $\Delta_{B}(T)$ are presented below.
\begin{figure}[ht]
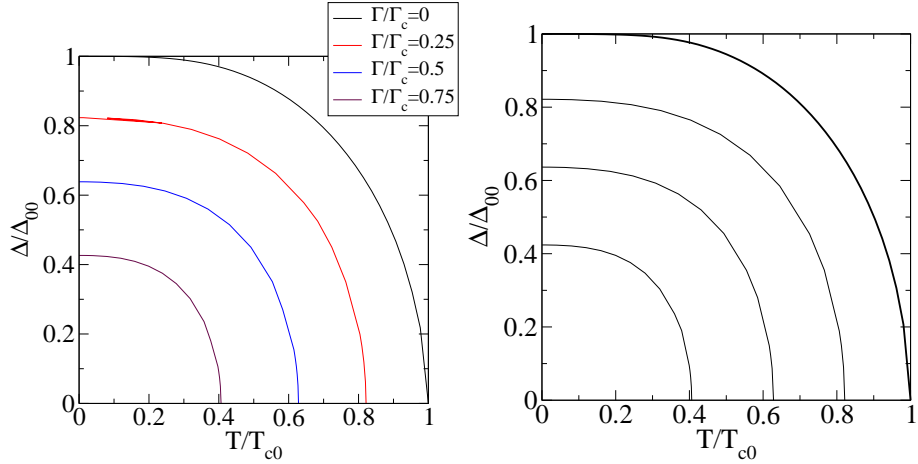

\includegraphics[width=6cm]{Adeltaimp.eps}
\includegraphics[width=6cm]{Bdeltaimp.eps}
\caption{The temperature dependent order parameters $\Delta_{A}(T)$ (left)
and $\Delta_{B}(T)$, for $\Gamma/\Gamma_{c} = 0, 0.25, 0.5$ and $0.75$.}
\end{figure}

Once $\Delta(\Gamma,T)$ has been determined, the thermodynamics of the system
can be analyzed.  We begin with the entropy:
\bea
S_{s} &=& -4\int_{0}^{\infty} dE \,N(E)(f\ln f+(1-f)\ln(1-f))
\eea
Here $f$ is the Fermi-Dirac distribution $(1+e^{\beta E})^{-1}$ with $\beta= 1/k_{B}T$ and N(E) is the quasiparticle density of states.  From the entropy
the electronic specific heat is determined via
\bea
C_{s} = T\frac{\partial S}{\partial T}
\eea
We show above the specific heat for both phases for all temperatures
below T$_{c}$ for a few impurity scattering rates.
\begin{figure}[ht]
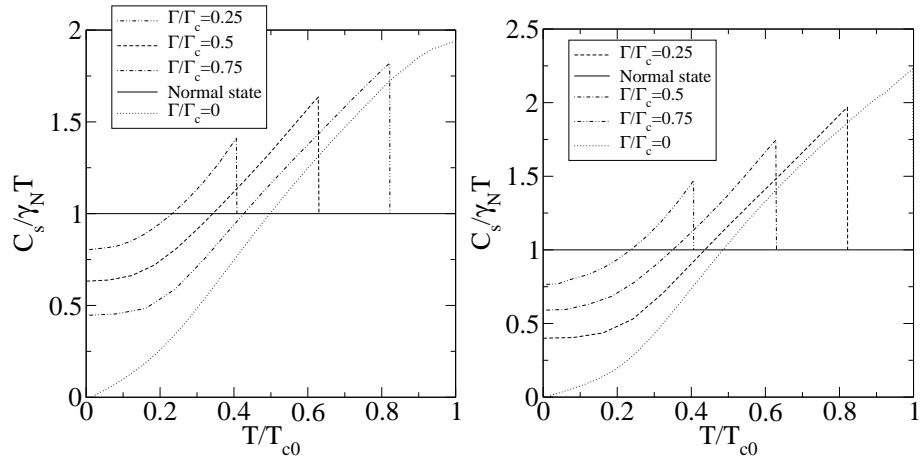

\includegraphics[width=6cm]{A_C_imp.eps}
\includegraphics[width=6cm]{B_C_imp.eps}
\caption{The specific heat for the A phase (left) and B phase is shown, for
impurity scattering rates $\Gamma/\Gamma_{c} = 0, 0.25, 0.5$ and $0.75$.}
\end{figure}
As expected, the specific heat is T-linear for small T, with the coefficient
equal to the zero-energy DOS.  The specific heat jump $\Delta C$ at 
T$_{c}$ decreases rapidly with impurity concentration for both phases.

The superfluid density is calculated as follows:
\bea
\rho_{s\,ab}(\Gamma,T) &=& 2 \pi T \sum_{0}^{\infty}\left\langle 
\frac{3}{2} \sin^{2} \theta
\frac{\Delta^{2}|f|^{2}}{({\tilde{\omega}_{n}}^{2}+\Delta^{2}|f|^{2})^{3/2}}
\right\rangle \\
\rho_{s\,c}(\Gamma,T) &=& 2 \pi T \sum_{0}^{\infty}\left\langle 3 \cos^{2} \theta
\frac{\Delta^{2}|f|^{2}}{({\tilde{\omega}_{n}}^{2}+\Delta^{2}|f|^{2})^{3/2}}
\right\rangle
\eea
Note that the superfluid density is isotropic for the cubic-symmetry
retaining A-phase, but is not for the symmetry-breaking B-phase.  We show 
below the superfluid density for both phases for a range of impurity
concentrations.  
\begin{figure}[ht!]
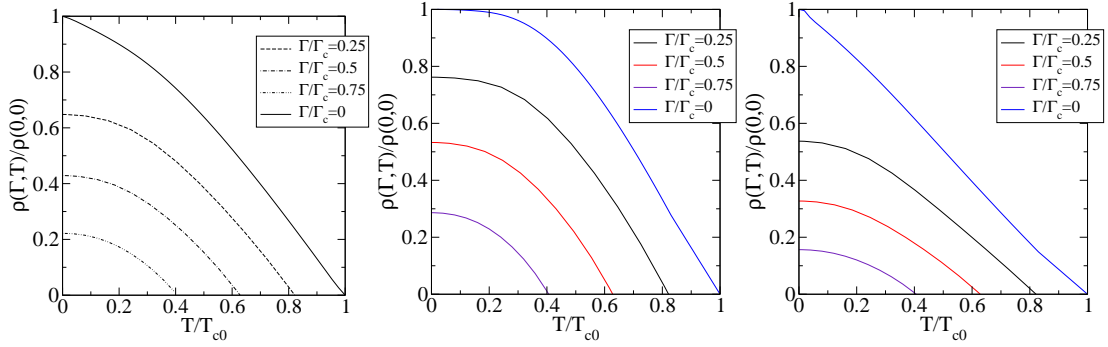

\includegraphics[width=4.8cm]{A_fluid.eps}
\includegraphics[width=4.8cm]{B_fluid_ab.eps}
\includegraphics[width=4.8cm]{B_fluid_c.eps}
\caption{From top left, the superfluid density for the A phase, for
the B phase ab plane, and for the B phase c axis.  Impurity scattering rates
are $\Gamma/\Gamma_{c} = 0, 0.25, 0.5$ and $0.75$, with lower impurity
scattering rates at the top of each plot.  Note that the top left and bottom
plots resemble that of d-wave superconductivity.}
\end{figure}
We observe that while the pure limit superfluid density varies linearly with
temperature at low temperature for both the A phase and the c-axis B phase
results, this dependence changes to a $T^{2}$ dependence when the effect of
impurities is considered. 

Finally, we consider the nuclear spin lattice relaxation rate $T_1^{-1}$, which
is related to the density-of-states N(E) by the following relation:
\bea
(T_{1} T)^{-1}/(T_{1} T)^{-1}_|T=T_{c} = \int_{0}^{\infty} dE N^{2}(E)
\mathrm{sech}^{2}(E/(2T))/2T
\eea
Plots of this quantity for both phases, for both the pure and impurity
scattering cases are shown.  In the zero energy limit 
\begin{figure}[h!]
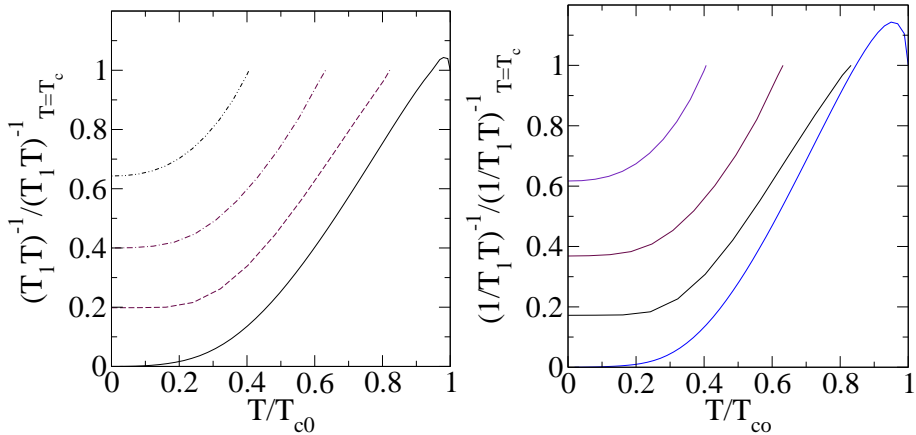

\includegraphics[width=6cm]{Arelax.eps}
\includegraphics[width=6cm]{Brelax.eps}
\caption{Plots of the nuclear spin lattice relaxation rate T$_{1}^{1}$ for the
A phase (left) and B phase are shown.   Impurity scattering rates
are $\Gamma/\Gamma_{c} = 0, 0.25, 0.5$ and $0.75$, with lower impurity
scattering rates at the top of each plot.}
\end{figure}
$(T_{1} T)^{-1}/(T_{1} T)^{-1}_{|T=T_{c}}$ becomes $N(0)^{2}$.  
We also note the presence of a small peak in 
both pure case results just below T$_{c}$.  As will be shown elsewhere 
\cite{parker}, this is a general feature of pure-case 
unconventional superconductors obeying BCS weak-coupling theory.  It results
from the distribution of N(E) away from the constant DOS observed in the
normal state.  Note that even the A-phase, for which the density-of-states
does not show the usual divergence at $E=\Delta$, shows a small peak.  The B
phase peak is much larger due to the increased density-of-states near 
$E=\Delta$; this divergence, in fact, is $\propto (E-\Delta)^{-1/4}$ and is
hence a somewhat stronger divergence than the logarithmic divergence
found in d-wave superconductivity.

\section{Conclusion}

We have calculated the effect of impurity scattering on the superconductivity
in PrOs$_{4}$Sb$_{12}$ within the unitary limit.  We find that 
the superconductivity is expected to be quite sensitive to impurity
concentration within this limit.

{\bf Acknowledgements}

We thank H. Won, K. Izawa, Y. Matsuda and H. Tou for many useful discussions.
DP and SH were supported by the Petroleum Research Fund.

\end{document}